\documentclass[twocolumn,preprintnumbers,prd,superscriptaddress,nofootinbib,floatfix,showpacs,showkeys]{revtex4-1}

\usepackage{float}
\usepackage{graphicx}

\usepackage{multirow}
\usepackage{natbib}
\usepackage{pifont}
\usepackage{amssymb}
\usepackage{bm}
\usepackage{amsfonts}
\usepackage{amsmath}
\usepackage{amssymb}
\usepackage[titletoc]{appendix}
\usepackage{color}
\usepackage{hyperref}
\usepackage{cleveref}
\usepackage[english]{babel}

\usepackage{booktabs}
\usepackage{mathtools}
\usepackage{subfigure}
\usepackage{enumerate}

\usepackage{dcolumn}

\usepackage{array}
\usepackage{ctable}
\usepackage{multirow}
\usepackage{siunitx}
\usepackage{longtable}
\usepackage{tabularx}
\usepackage{booktabs}

\usepackage{amsthm}
\usepackage{mathrsfs}

\usepackage{epsfig}
\usepackage{amscd}

\usepackage{bm}
\usepackage{natbib}
\usepackage{url}
\usepackage{xspace}
\usepackage{kantlipsum}

\graphicspath{{Graphics/}}

\def\be{\begin{equation}}
\def\ee{\end{equation}}
\def\bea{\begin{eqnarray}}
\def\eea{\end{eqnarray}}

\definecolor{vividviolet}{rgb}{0.62, 0.0, 1.0}
\definecolor{amaranth}{rgb}{0.9, 0.17, 0.31}
\definecolor{palatinateblue}{rgb}{0.15, 0.23, 0.89}
\definecolor{brightpink}{rgb}{1.0, 0.0, 0.5}
\definecolor{cornflowerblue}{rgb}{0.39, 0.58, 0.93}
\definecolor{deepcarminepink}{rgb}{0.94, 0.19, 0.22}
\definecolor{radicalred}{rgb}{1.0, 0.21, 0.37}

\hypersetup{ linktoc=all,
    colorlinks, linkcolor={palatinateblue},
    citecolor={brightpink}, urlcolor={amaranth}
}

\begin{document}

\title{Gravitational repulsive effects in 3D regular black holes}

\author{Orlando~Luongo}
\email{orlando.luongo@unicam.it}
\affiliation{Universit\`a di Camerino, Via Madonna delle Carceri 9, 62032 Camerino, Italy.}
\affiliation{SUNY Polytechnic Institute, 13502 Utica, New York, USA.}
\affiliation{INAF - Osservatorio Astronomico di Brera, Milano, Italy.}
\affiliation{Istituto Nazionale di Fisica Nucleare, Sezione di Perugia, 06123, Perugia,  Italy.}
\affiliation{Al-Farabi Kazakh National University, Al-Farabi av. 71, 050040 Almaty, Kazakhstan.}

\author{Hernando Quevedo}
\email{quevedo@nucleares.unam.mx}
\affiliation{Instituto de Ciencias Nucleares, Universidad Nacional Aut\'onoma de M\'exico, Mexico.}
\affiliation{Dipartimento di Fisica and Icra, Universit\`a di Roma “La Sapienza”, Roma, Italy.}

\author{S. N. Sajadi}
\email{naseh.sajadi@gmail.com}
\affiliation{School of Physics, Institute for Research in Fundamental Sciences (IPM), P. O. Box 19395-5531, Tehran, Iran.}

\begin{abstract}
    In this work, we consider the effects of repulsive gravity in an invariant way for four static 3D regular black holes, using the eigenvalues of the Riemann curvature tensor, the Ricci scalar, and the strong energy conditions. The eigenvalues of the solutions are non-vanishing asymptotically (in asymptotically AdS) and increase  as the source of gravity is approached, providing a radius at which the passage from attractive to repulsive gravity might occur.
    We compute the onsets and the regions of repulsive gravity  and conclude that the regular behavior of the solutions at the origin of coordinates can be interpreted as due to the presence of repulsive gravity, which also turns out to be related with the violation of the strong energy condition.
        We showed that in all of the solutions for the allowed region of parameters, gravity changes its sign, but the repulsive regions only for the non-logarithmic solution are affected by the mass that generates the regular black hole. The repulsive regions for the logarithmic solutions are dependent on electric charge and the AdS$_{3}$ length.
    The implications and physical consequences of these results are discussed in detail.
     \end{abstract}
\maketitle
\tableofcontents

\section{Introduction}

Einstein's general theory of relativity is the foundational framework to describe the large-scale universe and to predict the existence of a specific category of spacetime singularities, commonly referred to as \emph{black holes} \cite{Townsend:1997ku, Grumiller:2022qhx}. Recent empirical observations have provided direct confirmation of the existence of these enigmatic cosmic entities \cite{LIGOScientific:2016aoc, LIGOScientific:2018dkp, LIGOScientific:2019fpa, LIGOScientific:2019obb, EventHorizonTelescope:2019dse, EventHorizonTelescope:2019uob}. Moreover, within the realm of general relativity, theoretical investigations have indicated that, in addition to black hole solutions, other kinds of singularities might also exist \cite{Penrose:1964wq, Hawking:1966sx, Hawking:1970zqf, Ellis1979}.

A remarkable class of these solutions have the notable characteristic that the parameters associated with black holes violate the physical conditions to form an event horizon\footnote{Interestingly, despite extensive investigations, no scenarios of gravitational collapse have conclusively confirmed the validity of the cosmic censorship hypothesis, which posits that singularities are always concealed from observation. Consequently, there remains the possibility that, under specific circumstances, naked singularities may emerge during the course of mass distribution collapsing gravitationally \cite{Joshi:2008zz}. } \cite{Rpenros,Wald:1984rg}.

A prominent characteristic that arises in both naked and black hole singularities is the theoretical evidence for regions of \emph{repulsive gravity} and, more broadly, for repulsive effects, that represent well-consolidate theoretical phenomena for Schwarzschild, Kerr, and Kerr-Newman spacetimes \cite{Luongo:2014qoa} and so on, see e.g. \cite{Luongo:2023aib}. Furthermore, recent endeavors have focused on exploring the physical implications of potential regions with repulsive gravity within the framework of the homogeneous and isotropic Friedmann-Lemaître-Robertson-Walker spacetime \cite{Luongo:2015zaa}. This exploration seeks to address the perplexing issue of the observed cosmic acceleration \cite{Luongo:2015zaa}.

It is therefore natural to extend the hypothesis of having repulsive regions of gravity up to more complicated black hole solutions and/or exotic objects predicted from general relativity. In this respect, with the aim of circumventing the issue of singularities, certain modifications to classical gravity might be expected. Indeed, as a first attempt, one can eliminate the central singularity and the associated energy divergence linked to the electromagnetic field of a point charge \cite{Born:1934gh}. Consequently, black hole singularities could be replaced by regular regions filled with a specific form of matter or a false vacuum configuration that violates the strong energy condition. These objects, recently acquiring great relevance, are named \emph{regular black holes} \cite{bardeen1968, Ayon-Beato:1999kuh, Ayon-Beato:1998hmi, Dymnikova:1992ux, Dymnikova:2004zc, Novello:2000km, Bronnikov:2000vy, Burinskii:2002pz, Sajadi:2017glu, Hendi:2020knv, Torres:2022twv, Bronnikov:2022ofk, Karakasis:2023hni, Estrada:2023cyx, Kumar:2023ijg, Contreras:2017eza, Berej:2006cc, Elizalde:2002yz, Bronnikov:2000yz}.  One of the earliest examples of a regular black hole solution was put forth in Ref. \cite{bardeen1968}, whose formal structure can be argued within the framework of non-linear electrodynamics, exhibiting a magnetic monopole \cite{Ayon-Beato:1999kuh, Ayon-Beato:1998hmi}, capable of removing the singularity near the center.

Again, the repulsive effects within the framework of regular black holes have been explored \cite{Luongo:2023aib}. It has been demonstrated that standard black holes can replicate the outcomes of regular solutions under specific combinations of constants. Consequently, regular solutions might appear to be \emph{less predictive} than standard black holes, as there could exist a black hole whose repulsive effects can effectively degenerate into those of a regular solution through a specific combination of the free parameters characterizing the black hole itself.

Interestingly, due to the absence of local degrees of freedom, general relativity appears simpler in three than in four spacetime dimensions\footnote{Including local degrees of freedom makes the theory closer to its four-dimensional counterpart; this can be achieved by deforming the theory, giving a mass to the graviton \cite{Deser:1982vy, Bergshoeff:2009hq}.}. In spite of this simplicity, the theory does contain black holes when a negative cosmological constant is included\footnote{In 3D flat spacetime, the only solution with the horizon is flat space cosmology \cite{Bagchi:2013lma, Setare:2020mej}. The other solutions are the kink-like solution with the gravitational field which is described by a conical space with its deficit angle corresponding to the mass of the particle. \cite{Banados:1992wn, Carlip:1995qv}.}. This black hole spacetime is obtained by identifying certain points of the anti-de Sitter space which is characterized by the mass, angular momentum, and cosmological constant. Hence, motivated by the \emph{BTZ black hole}, the extended 3D black hole solutions have received much attention from a theoretical standpoint in recent years\footnote{In this direction, one can find electrically charged black holes, dilatonic black holes, black holes arising in string theory, black holes in
topologically massive gravity, and warped-AdS black holes \cite{Carlip:1995qv, Rincon:2018sgd, Panotopoulos:2018pvu, Podolsky:2018zha, Sharif:2019mzv, Ali:2022zox, Priyadarshinee:2023cmi, Karakasis:2023ljt}.} and, so, in lower-dimensional spacetimes, smooth regular black hole solutions have also been proposed.

For example, in (2+1)-dimensional spacetime, a regular black hole is constructed by introducing non-linear electrodynamics as a source for such a configuration \cite{He:2017ujy, Bueno:2021krl, Aros:2019quj, Hendi:2022opt, Cataldo:2000ns}. Numerous intriguing properties of both static and rotating regular black holes have been thoroughly investigated within the scientific literature. These studies have encompassed various aspects, including thermodynamics, dynamical stability, geodesics, and more. For an in-depth examination of such regular black holes \cite{Nicolini:2023hub, Kumar:2022fqo, Banerjee:2022bxg, Banerjee:2022iok, Villani:2021lmo, Kumar:2020ltt, Nam:2019zyk, Flachi:2012nv}. The classification of the solutions in 3D according to cotton tensor scalars has been done in \cite{Podolsky:2023qiu}.

Motivated by these findings, we wonder whether repulsive regions of gravity may change if one employs (2+1)-dimensional black hole solutions. To do so, we consider an invariant treatment in which first-order curvature invariants are used. We then assume as background four solutions that resemble the structure of Bardeen and Hayward spacetimes. In all the metrics here involved, we assumed the non-linear electrodynamics contribution, made smoother in the center, and behaving differently at larger radii, i.e., typically involving an anti-de Sitter non-vanishing phase. The regions of repulsion are found by analyzing the behavior of the corresponding curvature eigenvalues. The structures of solutions, albeit different from the four-dimensional cases, behave quite similar to standard solutions. Phrasing it differently, we first find regions of repulsive gravity that arise analogously to other solutions and show how repulsive gravity occurs. A corresponding physical interpretation of those regions is also reported. A comparison with previous literature is finally proposed.

The paper is organized as follows. In Sect. \ref{sezione2}, we introduce the basic demands of repulsive gravity through curvature invariants for a given metric. In Sect. \ref{sezione3}, we explore four static regular black holes in three dimensions and calculate for each of them the main geometric features related to the curvature invariants. A comprehensive discussion of our findings is reported in Sect. \ref{sezione4}, where we interpret our outcomes and compare them with previous solutions. Finally, in Sect. \ref{sezione5}, we report our conclusions and perspectives of our work.

\section{Curvature eigenvalues}\label{sezione2}

A change in the gravity sign can be easily described by assuming a change in the mass sign. Conversely, if we consider a negative mass, it becomes clear that the sign of gravity would also change accordingly, albeit negative masses are clearly nonphysical. Instead, if one takes the Ricci scalar, being proportional to the trace of the energy-momentum tensor, inverting its sign would naively invert the sign of the mass and/or the pressure as well. Unfortunately, this naive approach does not properly work when dealing with vacuum solutions of Einstein's field equations. Hence, to address this issue, curvature invariants can be utilized. However, quadratic invariants, such as the Kretschmann scalar, contain only the square of the mass so that a change of sign is imperceptible. To overcome this limitation, it was proposed to use the curvature eigenvalues, which are first-order invariants \cite{quev12,Luongo:2014qoa}.

To compute the curvature eigenvalues, it is convenient to utilize the bivector representation of the Riemann tensor, which is based upon the use of a local orthonormal tetrad, $e^a$, defined by means of
\begin{equation}
ds^2=g_{\mu\nu} dx^\mu \otimes dx^\nu = \eta_{a b}e^{a}\otimes e^{b}
\end{equation}
with $\eta_{ab}=diag(-1,1,1)$. The connection one-form, $\omega^a_{\ b}$ is determined by the first Cartan structure equation
\begin{equation}\label{eqq2}
de^{a}+\omega^{a}{}_{b}\wedge e^{b}=0,
\end{equation}
whereas the curvature two-form is
\begin{equation}\label{eqq3q}
\Omega^{a}{}_{b}=d\omega^{a}{}_{b}+\omega^{a}{}_{c}\wedge\omega^{c}{}_{b}=\dfrac{1}{2}R^{a}{}_{b c d}e^{c}\wedge e^{d}.
\end{equation}
They allow us to compute the components of the Riemann curvature tensor in the local orthonormal frame.

The bivector representation is a matrix representation of the Riemann tensor that
can easily be obtained by introducing
the bivector index $A = 1,2,3$, which corresponds to two tetrad indices, $A\to ab$. To this end, we choose
to the following convention
\begin{equation}
1\to 12,\;\;\;\;2\to 13,\;\;\;3\to 23 \ ,
\label{conv}
\end{equation}
leading to the representation of the Riemann curvature tensor $R_{abcd}$ as the ($3\times 3$)-matrix $R _{AB} $.

For the particular solutions we will investigate in this work, the spacetime geometry in circularly symmetric 3D in coordinates ($t, r, \phi$) can be expressed as
\begin{equation}\label{eq3dmetric}
ds^2=-f(r)dt^2+\dfrac{dr^2}{f(r)}+r^{2}d\phi^2\;,
\end{equation}
where the Schwarzschild-like coordinates, here used, are real, having $t,r\in ]-\infty,+\infty[$ and $\phi \in [0,2\pi]$.
We thus have
\begin{equation}
e^{1}=\sqrt{f}dt,\;\;\;e^{2}=\dfrac{dr}{\sqrt{f}},\;\;\;e^{3}=rd\phi.
\end{equation}
The connection forms are determined using Eq. \eqref{eqq2} as follows
\begin{equation}
\omega^{1}{}_{2}=\dfrac{f^{\prime}}{2\sqrt{f}}e^{1},\;\;\;\omega^{2}{}_{3}=-\dfrac{\sqrt{f}}{r}e^{3} \ .
\end{equation}
Then, using Eq. \eqref{eqq3q}, we obtain
\begin{equation}
R_{1212}=\dfrac{f^{\prime\prime}}{2},\;\;\;\;R_{1313}=\dfrac{f^{\prime}}{2r},\;\;\;R_{2323}=-\dfrac{f^{\prime}}{2r}.
\end{equation}

Using the convention (\ref{conv}), the matrix $R_{AB}$ can be written as
\begin{center}
\begin{equation*}
R_{AB}=  \begin{bmatrix}
\dfrac{f^{\prime\prime}}{2} & 0&0 \\
0 & \dfrac{f^{\prime}}{2r}&0\\
 0 & 0&-\dfrac{f^{\prime}}{2r}
 \end{bmatrix}.
\end{equation*}
\end{center}
Hence, the curvature eigenvalues, $\lambda_{i}$, are given by
\begin{equation}
\lambda_{1}=\dfrac{f^{\prime\prime}}{2},\;\;\;\lambda_{2}=\dfrac{f^{\prime}}{2r},\;\;\;\lambda_{3}=-\dfrac{f^{\prime}}{2r}.
\label{eigenv}
\end{equation}

As found in Ref. \cite{Luongo:2023aib}, the eigenvalues of the curvature tensor have the following two main properties:

\begin{itemize}
    \item[-] the change of sign of at least one eigenvalue indicates the transition to repulsive gravity,
    \item[-] The presence of a stationary point, i.e., maximum or minimum, in an eigenvalue indicates the repulsive gravity onset. The greatest stationary point is called the repulsion radius.
\end{itemize}

In the case of black holes in 4D, we expect the eigenvalues to vanish at infinity.
As we approach the black hole, in the presence of attractive gravity only, the eigenvalues will increase until they reach their maximum value near the curvature singularity. If repulsive gravity becomes dominant at a given point, one would expect at that point a change in the sign of at least one eigenvalue. This implies that the eigenvalue must have an extremum at some point before it changes its sign. According to this analysis, we define the radius of repulsion as the first extremum that appears in a curvature eigenvalues from the infinity, i.e.,
\begin{equation}
   \left. \dfrac{d\lambda_{i}}{dr}\right\vert_{r=r_{rep}}=0,
\end{equation}
where $r$ is a radial coordinate and $\lambda_{i}$ is any eigenvalue.

In the case of 3D, the situation is different. Since all the black holes in 3D are locally AdS$_{3}$ in the asymptotic regime, the gravitational field and eigenvalues at spatial infinity do not vanish. For example, the Riemann tensor for maximally symmetric spacetimes in $n$ dimensions shows a constant value for AdS metric that, for a maximally symmetric $g_{\mu\nu}$, reads
\begin{equation}
 R_{\mu\rho\nu\sigma}=\dfrac{2\Lambda}{(n-1)(n-2)}(g_{\mu\nu}g_{\rho\sigma}-g_{\mu\sigma}g_{\rho\nu}).
 \end{equation}

Furthermore, since we are dealing with regular spacetimes, we expect that all the eigenvalues will be free of singularities in the entire spacetime, tending to finite values near the gravitational source.

In the following, we search for these regions in 3D regular black holes.

\section{Static regular black holes}\label{sezione3}

The simplest black hole solution in 3D space-time is the BTZ spacetime, which is a solution of Einstein's gravity with cosmological constant \cite{Banados:1992wn}. When other fields exist in the action, the metric function may include other terms. If the extra term in the metric is a $\log$-term, then the source of the black hole is the electromagnetic field\footnote{It should be noted that the logarithmic solutions can also be solutions of 3D gravity theories at the chiral point.}. If the additional term is non-logarithmic, then the field added to the action is that of non-linear electrodynamics, whose asymptotic limit is not the Maxwell field. The action of 3D Einstein's gravity is given by
\begin{equation}
    \mathcal{A}=\int d^{3}x\sqrt{-g}\left(R-2\Lambda +\mathcal{L}_{m} \right),
\end{equation}
where $\Lambda=-1/\ell^2$ is the cosmological constant, and $\mathcal{L}_{m}$ is the Lagrangian of the additional field.

Accordingly, we considered two types of regular black holes and four different solutions, two of them contain a $\log$ contribution and are named Bardeen-like, whereas the other two solutions contain no $\log$ term and are named Hayward-like. The role of the additional term consists of freeing the solutions from singularities even at the centers located at  $r=0$.

\subsection{{Energy conditions}}
\label{sec:enc}

The strong energy condition states that for any time-like vector $\xi^{\mu}$ at any point of the spacetime the stress-energy tensor $T^{\mu}_{\nu}=diag\left(-\rho,P_{r},P_{t}\right)$ and its trace $T$ satisfy
\begin{equation}
    \left(T_{\mu \nu}-\dfrac{1}{2}g_{\mu\nu}T\right)\xi^{\mu}\xi^{\nu}\geq 0.
\end{equation}
In terms of the components of the energy-momentum tensor, it  is given as
\begin{equation}
  \text{for each $i$}:\;\;\;  \rho+P_{i}\geq 0,\;\;\;\text{and}\;\;\;\rho+\sum_{i} P_{i}\geq0.
\end{equation}
Explicitly, for the metric \eqref{eq3dmetric} we have
\begin{align}
\rho+P_{r}=&0,\label{cond15}\\
    \rho+P_{t}\geq&0\;\;\;\to\;\;\;f^{\prime\prime}-\dfrac{1}{r}f^{\prime}\geq0,\label{cond16}\\
    \rho+P_{r}+P_{t}\geq& 0,\;\;\;\to\;\;\;\dfrac{1}{2}f^{\prime\prime}-\dfrac{1}{\ell^{2}}\geq0.\label{cond17}
\end{align}
It should be noted that we have used $\rho=-(G^{1}_{1}-1/\ell^2)$ and $P_{i}=G^{i}_{i}-1/\ell^2$.
The condition \eqref{cond15} is identically satisfied. However, as we will see below, the strong energy condition \eqref{cond16} can be violated at a particular radius that we denote as
 $r_{sec}$.

Below, we summarize our findings focusing on each solution in detail.

\subsection{The first solution: 3D anti-de Sitter with log correction and topological charge}

We consider first the regular solution \cite{He:2017ujy,Cataldo:2000ns}
\begin{equation}\label{eqf2}
f(r)=-M+\dfrac{r^2}{\ell^2}-q^{2}\ln\left(\dfrac{{q^2}+r^2}{\ell^2}\right),
\end{equation}
where $M$ and $q$ represent the mass and electric charge, respectively. The
 BTZ black hole is recovered when $q$ tends to zero. The metric function contains a   $\log$ term, indicating that it is a solution to Einstein's field equations coupled to non-linear electrodynamics with Lagrangian, $\mathcal{L}_{m}(r)=q^2(r^2-q^2)/(q^2+r^2)^2$. This particular case of non-linear electrodynamics exhibits the linear limit of Maxwell's theory in the weak field approximation and satisfies the weak energy condition.

The roots of Eq. \eqref{eqf2}, which correspond to the horizon radii, are given as
\begin{equation}
r_{\pm}=\pm\sqrt{\ell^2 e^{-\mathcal{W}\left(-\frac{e^{-\frac{q^2+M\ell^2}{q^2\ell^2}}}{q^2}\right)+\frac{q^2+M\ell^2}{q^2\ell^2}}-q^2},
\end{equation}
where $\mathcal{W}(x)$ is the \emph{LambertW function}, i.e., a multi-valued function satisfying
$\mathcal{W}(x)e^{\mathcal{W}(x)}=x$ for the independent variable $x$, which can be either real or complex \cite{lambert}.

According to Eq.(\ref{eigenv}),
the corresponding curvature eigenvalues are \begin{align}\label{soluzionimetrica1}
\lambda_{1}=&\dfrac{(1-\ell^2)q^4+r^2q^2(2+\ell^2)+r^4}{\ell^2(q^2+r^2)^{2}},\\
\lambda_{2}=&-\lambda_{3}=\dfrac{q^2(1-\ell^2)+r^2}{\ell^2(q^2+r^2)}.
\end{align}
For $r\to \infty$, then $\lambda_{1}=\lambda_{2}=-\lambda_{3}=1/\ell^2$.
When approaching the source from infinity, the first extremum is located in $\lambda_{1}$ at
\begin{equation}
    r_{rep}=\sqrt{3}q ,
\label{rep}
\end{equation}
which determines the place of the repulsion onset.
The event horizon radius coincides with the repulsion radius at $r_{+}=r_{rep}=\sqrt{\frac{-3M}{\mathcal{W}\left(-\frac{4M\exp{(-3/\ell^2)}}{\ell^2}\right)}}$ or $M/q^2=3/\ell^2-2\ln{2q/\ell}$.
Furthermore, it is easy to see that $\lambda_{1}$ and $\lambda_{2}$ become zero at
\begin{align}
r^{(1)}_{dom} =&\dfrac{\sqrt{-4-2\ell^2+2\sqrt{\ell^4+8\ell^2}}q}{2},\\
r^{(2)}_{dom} =&q\sqrt{-1+\ell^2}.
\end{align}
According to our definition, we take the greatest value which indicates the region where repulsion dominates (the first zero from infinity), i.e. $r^{(2)}_{dom}$.

The strong energy condition (\ref{cond16}) turns out to be identically satisfied for $r\geq 0$. However, condition (\ref{cond17}) is violated  at the radius
\begin{equation}
    r_{sec} = q .
\end{equation}
  Notice the interesting relationship $r_{sec}= r_{rep}/\sqrt{3}$, indicating that the repulsion radius can be associated with the location, where the strong energy condition is violated. Moreover, notice that $r_{dom}^{(1)}$ and $r_{dom}^{(2)}$ are both functions of $l$ and $q$, in contrast to $r_{rep}$ and $r_{sec}$, which depend only on $q$.

\begin{figure*}[ht!]
	\centering
	\subfigure[]{\includegraphics[width=0.8\columnwidth]{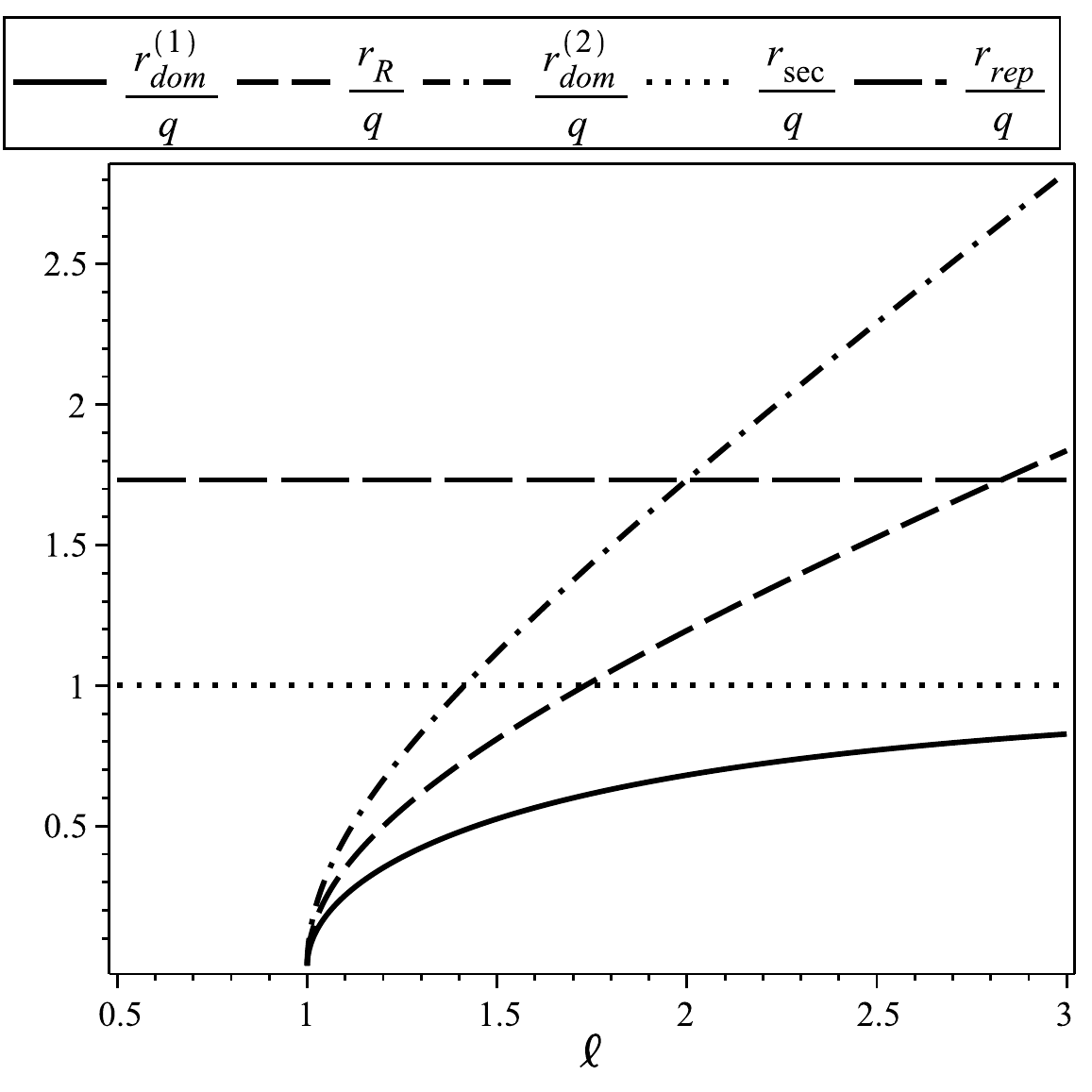}}~~~~~~~~~~
	\subfigure[]{\includegraphics[width=0.8\columnwidth]{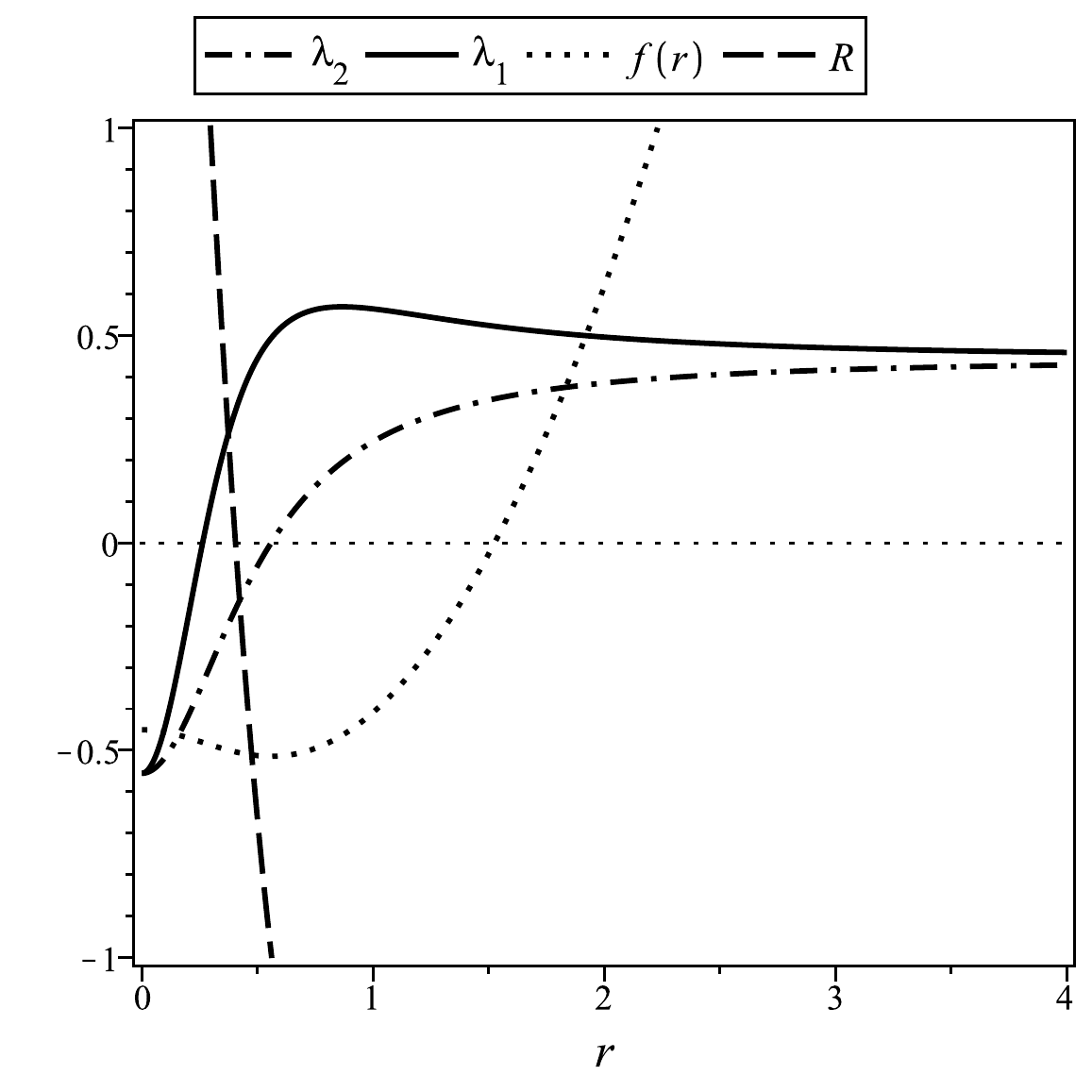}}
	\caption{Plots of ${r^{(2)}_{dom}/q}$, $r_{R}/q$, ${r^{(1)}_{dom}/q}$, $r_{rep}/q$ and $r_{sec}/q$ in terms of $\ell$ have been shown (left). Plots of the eigenvalues, ${\lambda_{1}}$, ${\lambda_{2}}$, ${f(r)}$, and $R$ for solution \eqref{eqf2} for $\ell=1.5,M=1,q=0.5$ (right).
 }
 	\label{fig1}
\end{figure*}

As can be seen from Fig.\ref{fig1}(a), for a constant value of $\ell$, $r^{(2)}_{dom}/q$ is greater than $r^{(1)}_{dom}/q$ and $r_{R}/q$. For $\ell=\sqrt{2}$, the $r^{(2)}_{dom}=r_{sec}$ and for $\ell=\sqrt{3}$ the $r_{R}=r_{sec}$.
For $\ell=1$, all radii are equal. For $\ell\leq 2$, the $r^{(2)}_{dom}\leq r_{rep} $. Therefore the allowed possible region for $\ell$ is $1\leq\ell\leq 2$.

In Fig. \ref{fig1}(b), we show the behavior of the eigenvalues, which, as expected, are free of singularities everywhere and tend to a finite value at the origin of coordinates.
The main result is that repulsive gravity is present in both eigenvalues and, in fact, it becomes the dominant component as the origin of coordinates is approached, forcing the eigenvalues to become finite at $r=0$. Thus, we conclude that the regular behavior of this black hole is a consequence of the presence of repulsive gravity. Another important and unexpected result is that the eigenvalues contain information about the energy conditions. In fact, it can be shown that {$r_{sec_2}=r_{rep}/\sqrt{3}$} determines the location where the strong energy condition is violated.

\begin{figure}
    \centering
    \includegraphics[scale=0.35]{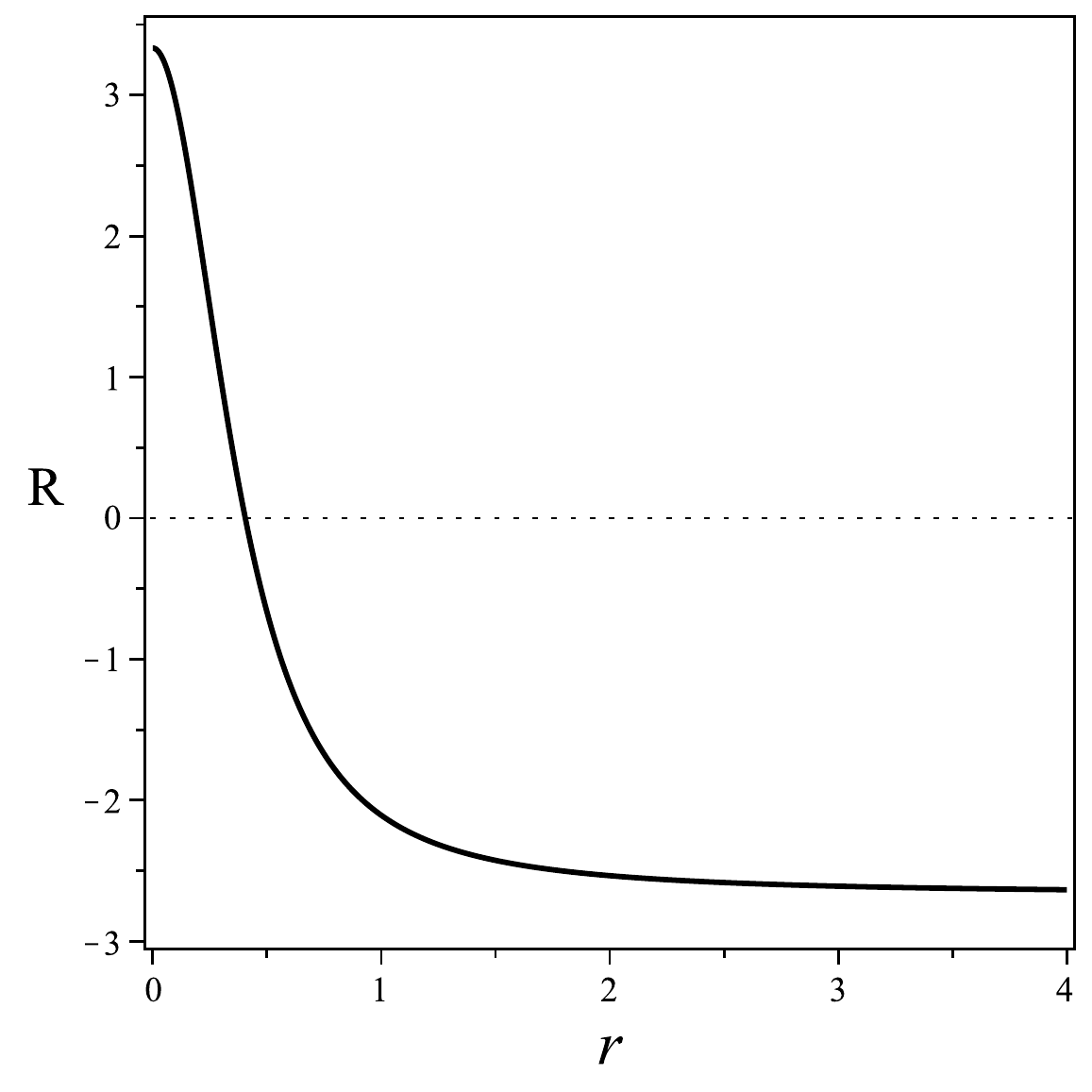}
    \caption{The Ricci scalar as a function of the radial coordinate for $\ell=1.5, M=1,q=0.5$ }
    \label{fig1c}
\end{figure}

As mentioned above, in 3D gravity the Ricci scalar is not trivial. Therefore, we can compare its behavior with that of the eigenvalues. A straightforward computation shows that the Ricci scalar reads
\begin{equation}
R=\dfrac{2(-3q^4-6q^2r^2-3r^4+3q^4\ell^2+q^2r^2\ell^2)}{\ell^2(q^2+r^2)^2}.
\end{equation}
Furthermore, one can see that it  changes its sign at the radius
\begin{equation}
r_{R}=\dfrac{\sqrt{-36+6\ell^2+6\sqrt{\ell^4+24\ell^2}}q}{6}.
\end{equation}
In Fig. \ref{fig1c}, we show the behavior of $R$ as a function of the radial distance $r$. We see that the Ricci scalar is free of singularities and tends to a finite value at the origin of coordinates.

The Ricci scalar also shows the presence of repulsive gravity with a region of repulsion dominance for $r<r_R$ and a finite value at $r=0$. We conclude that the Ricci scalar corroborates the results we have found at the level of the eigenvalues. However, we could not find any physical interpretation for the radius $r_R$, which is always located between the radius of the dominance of repulsion, $r=r_{dom }^{(2)}$, and, $r=r_{dom}^{(1)}$ (see Fig. \ref{fig1}(b)).

\subsection{The second solution:
3D anti-de-Sitter with a Hayward-like correction}

We now consider the solution  \cite{Aros:2019quj},\cite{Hendi:2022opt}
\begin{equation}\label{eqq10}
f(r)=1+\dfrac{r^2}{\ell^2}-\dfrac{8Mr^2}{2\zeta M+r^2} \ ,
\end{equation}
where $M$ and $\zeta$ are two real integration constants. This regular black hole is a solution in  Einstein gravity coupled to non-linear electrodynamics with Lagrangian $\mathcal{L}_{m}(r)=-(2M^2\zeta(2M\zeta-3 r^2))/\pi(2M\zeta+r^2)^{3}$. In this case, the non-linear electrodynamics does not contain the Maxwell theory limit in the weak field approximation.
For $\zeta=0$ the metric (\ref{eqq10}) reduces to the BTZ black hole and for $\zeta=M=0$ it corresponds to the $AdS_{3}$ spacetime. For a certain range of the parameters $M$ and $\zeta$, the solution describes a black hole with Killing horizons located at $r_{\pm}$ such that $f(r_{\pm})=0$, namely,
\begin{widetext}
\begin{equation}
r_{\pm}=\dfrac{1}{2}\sqrt{16M\ell^2-4\zeta M-2\ell^2\pm2\sqrt{\ell^4-4\zeta M\ell^2 -16M\ell^4+4M^2\zeta^2-32M^2\zeta\ell^2+64M^2\ell^4}}.
\end{equation}
\end{widetext}
For $\ell^4-4\zeta M\ell^2 -16M\ell^4+4M^2\zeta^2-32M^2\zeta\ell^2+64M^2\ell^4=0$, we have $r_{+}=r_{-}$, which corresponds to extremal black holes.

\begin{figure*}[ht!]
	\centering
	\subfigure[]{\includegraphics[width=0.8\columnwidth]{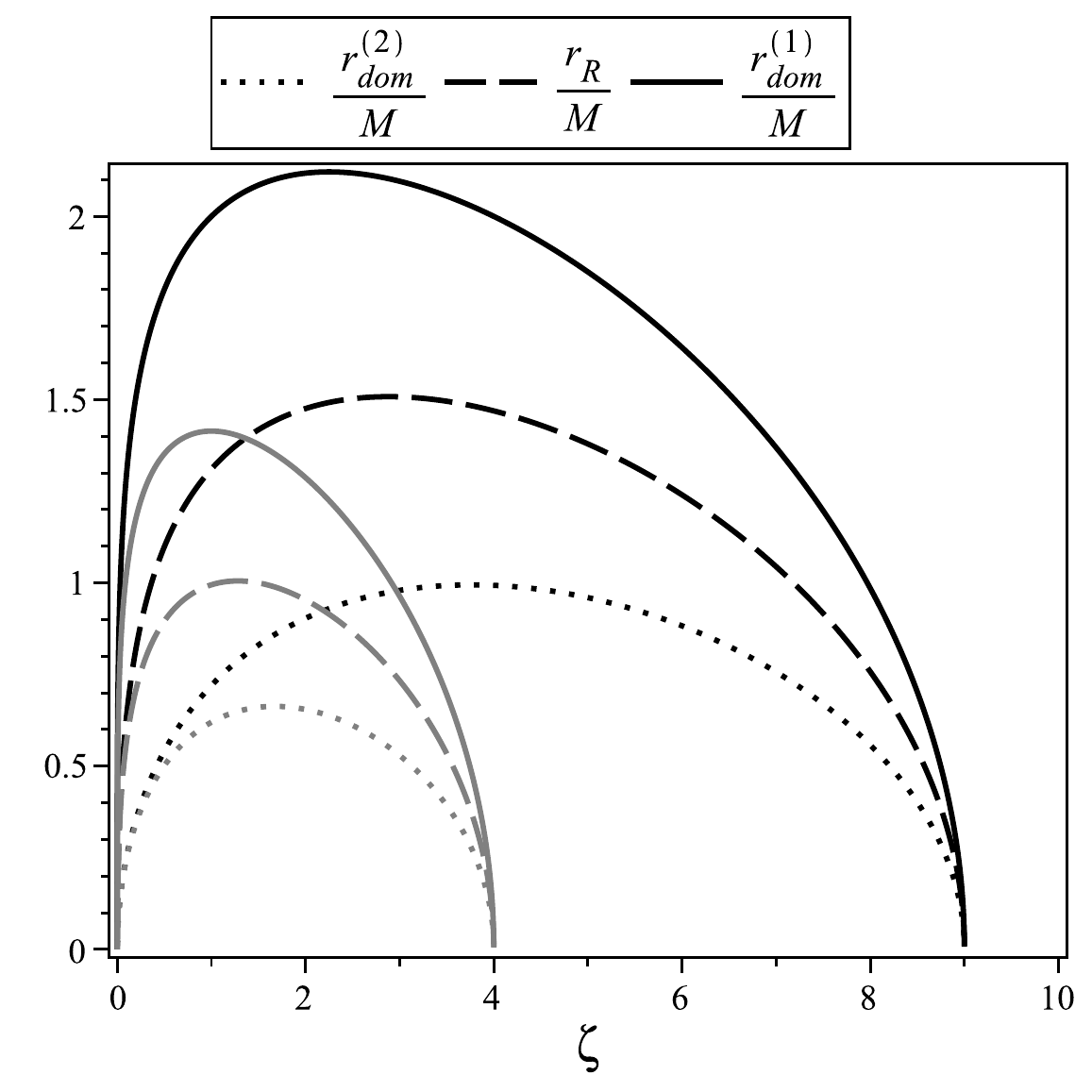}}~~~~~~~~~~
	\subfigure[]{\includegraphics[width=0.8\columnwidth]{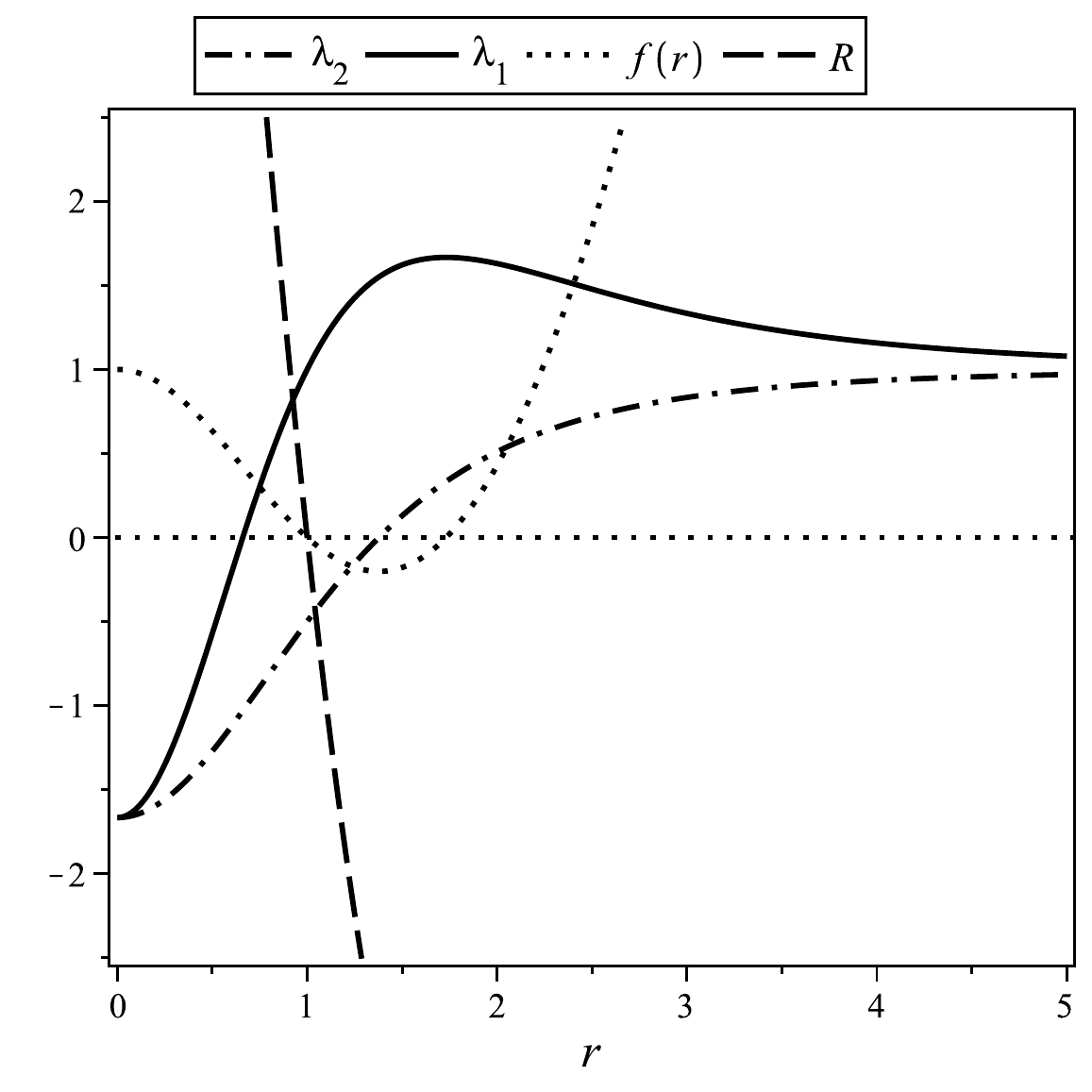}}
	\caption{Plots of ${r^{(1)}_{dom}/M}$, $r_{R}/M$ and ${r^{(2)}_{dom}/M}$ in terms of $\zeta$ for $\ell=\textcolor{gray}{1},1.5$ have been shown (left). Plots of the eigenvalues, ${\lambda_{1}}$, ${\lambda_{2}}$ and ${f(r)}$, $R$ for the second solution for $\ell=1,M=1,\zeta=1.5$. }
	\label{fig2}
\end{figure*}

Computing the curvature eigenvalues from Eq.(\ref{eigenv}), we obtain
 \begin{align}\label{soluzionimodello2}
 \lambda_{1}=&\dfrac{r^6+6\zeta Mr^{4}+12(\zeta+4\ell^2)\zeta r^2 M^2+8\zeta^2 M^3(\zeta-32\ell^2)}{\ell^2(2\zeta M+r^2)^{3}},\\
 \lambda_{2}=&-\lambda_{3}=\dfrac{r^4+4\zeta M r^{2}+4\zeta M^{2}(\zeta-4\ell^2)}{\ell^{2}(2\zeta M+r^{2})^{2}}.
  \end{align}
In the limit $r\to\infty$, the eigenvalues $\lambda_{i}$ become $1/\ell^2$.
The first extremum that is reached when approaching from infinity is located at $r_{rep}=\sqrt{2\zeta M}$, which corresponds to a local maximum of $\lambda_{1}$. The event horizon radius coincides with repulsive radius at $r_{+}=r_{rep}=\sqrt{-1+4M}\ell$ or $M=\ell^2/(2(2\ell^2-\zeta))$.
Furthermore, it is easy to see that $\lambda_{1}$ and $\lambda_{2}$ become zero at
\begin{align}\label{repulsione2}
r^{(1)}_{dom} =&\sqrt{4M\ell\sqrt{\zeta}-2\zeta M},\\
r^{(2)}_{dom} =&\dfrac{\zeta^{\frac{1}{4}}\sqrt{2M(2(\ell X)^{\frac{2}{3}}-2\ell^{\frac{4}{3}}-\sqrt{\zeta}X^{\frac{1}{3}})}}{X^{\frac{1}{6}}},
\end{align}
where $X=\zeta^{\frac{1}{2}}+(\ell^2+\zeta)^{\frac{1}{2}}$.
According to our definition, we take the largest value, which indicates the region where repulsion dominates (the first zero from infinity), i.e., $r^{(1)}_{dom}$. Further, the radius at which the strong energy condition is violated is equal to $r_{sec}=r_{rep}/\sqrt{3}$. For $\zeta\geq\ell^2$, we have $r^{(1)}_{dom}\leq r_{rep}$.
Moreover, the Ricci scalar reads
\begin{widetext}
\begin{equation}
R=\dfrac{2(-24\zeta^3 M^3-36\zeta^2 M^2r^2-18\zeta Mr^{4}-3r^{6}+96M^{3}\ell^2\zeta^2-16M^2\zeta\ell^2 r^{2})}{\ell^2(2\zeta M+r^2)^{3}}.
\end{equation}
\end{widetext}
The Ricci scalar changes sign at
\begin{align}
r_{R}=\dfrac{\sqrt{6MY(2Y^2-2\zeta \ell^2-3\zeta Y)}}{3Y},
\end{align}
with $Y=[9\ell^2\zeta^2+\sqrt{\ell^6\zeta^3+81\ell^4\zeta^4}]^{\frac{1}{3}}
$, exhibiting repulsive gravity at distinct domains. As it can be noticed from Fig. \eqref{fig2}a, for a constant value of $\ell$, $r^{(2)}_{dom}/q$ is larger than $r^{(1)}_{dom}/q$ and $r_{R}/q$. Further, for each value of $\ell$, $\zeta$ is restricted to $0\leq\zeta\leq 4\ell^{2}$. For $\zeta=0;4\ell^2$, all radii appear the same. For $\zeta\geq \ell^2$, the $r^{(2)}_{dom}\leq r_{rep} $, therefore the allowed region for $\zeta$ is $\ell^2\leq\zeta\leq 4\ell^2$.

From the above results we can again conclude that repulsive gravity can be considered as responsible for the regular behavior of this black hole. Moreover, we notice again the interesting relation between repulsive gravity and the violation of the strong energy condition.

\subsection{The third solution: 3D anti-de-Sitter with a Hayward-like correction}

The third regular black hole solution considered here is \cite{Bueno:2021krl}
\begin{equation}\label{eqmet30}
f(r)=\dfrac{-M+\dfrac{r^2}{\ell^2}+\dfrac{2\xi\ell^2}{r^2}}{1+\dfrac{2\xi\ell^2}{r^2}},
\end{equation}
where $M$ is a real integration constant, interpreted as mass, and $\xi$ is a real constant related to the coupling of the theory. This regular black hole is a solution to Einstein's gravity  non-minimally coupled to a scalar field as follows \cite{Bueno:2021krl}
\begin{equation}
    \mathcal{L}_{m}=-\ell^{2}[(\partial{\varphi})^{4}-\sqrt{\xi}(3R^{b c}\partial_{b}\varphi\partial_{c}\varphi-R(\partial\varphi)^2)],
\end{equation}
with $\varphi=\sqrt{2\sqrt{\xi}}\phi$. For $\xi=0$ the metric (\ref{eqq10}) reduces to the BTZ black hole.
The roots of Eq. \eqref{eqmet30} are given by
\begin{equation}
r_{\pm}=\dfrac{\ell}{2}\sqrt{2M\pm 2\sqrt{M^2-8\xi}}.
\end{equation}
For $M=2\sqrt{2\xi}$, the $r_{+}=r_{-}$.

According to Eq.(\ref{eigenv}), the curvature eigenvalues, $\lambda_{i}$, for this metric, can be written as
\begin{align}\label{soluzionimodello3}
\lambda_{1}=&\dfrac{r^6+6\xi\ell^2 r^4+6\xi r^2\ell^4(1+M+4\xi)-4\xi^2\ell^6(1+M)}{\ell^2(r^2+2\xi\ell^2)^{3}},\\
\lambda_{2}=&-\lambda_{3}=\dfrac{r^4+4\xi\ell^2r^2-2\xi(1+M)\ell^4}{\ell^2(r^2+2\xi\ell^2)^2}.
\end{align}
Asymptotically,  $r\to\infty$, the eigenvalues $\lambda_{i}$ tend to the de-Sitter value  $1/\ell^2$.
The first extremum that is reached when approaching from infinity is located at $r_{rep}=\sqrt{2\xi }\ell$, which corresponds to a local maximum of $\lambda_{1}$.
The event horizon radius in terms of the repulsion radius can be expressed as
\begin{equation}
    r_{+}=\dfrac{\ell\sqrt{2M}}{2}\sqrt{1+\sqrt{1-\dfrac{4r_{rep}^{2}}{\ell^2 M^2}}}.
\end{equation}
We see that a black hole can exist only if $0<r_{rep}<\ell M/2$.
The event horizon radius coincides with repulsive radius at $r_{+}=r_{rep}=\sqrt{-1+M}\ell$ or $M=2\xi+1$.
Furthermore, it is easy to see that $\lambda_{1}$ and $\lambda_{2}$ become zero at
\begin{align}\label{repulsione3}
r^{(1)}_{dom} =&\ell\sqrt{-2\xi+\sqrt{4\xi^2+2\xi+2\xi M}},\\
r^{(2)}_{dom} =&\ell\sqrt{1-2\xi-\Big[\dfrac{4\xi^2+2\xi-2\xi M}{X^{\frac{1}{3}}}\Big]},
\end{align}
here $X=16\xi^3 +8\xi^2 +8M\xi^2+2\xi(1+M+2\xi)\sqrt{2\xi(1+M+10\xi)}$. The first zero from infinity corresponds to the region where repulsion dominates, i.e. $r^{(1)}_{dom}$. Further, the place where the strong energy condition is violated is equal to $r_{sec}=r_{rep}/\sqrt{3}$. For $\xi\geq(1+M)/6$, we have $r^{(1)}_{dom}\leq r_{rep}$.
The Ricci scalar reads
\begin{widetext}
    \begin{equation}
R=\dfrac{2(-3r^6-18r^4\xi\ell^2 -40r^2\xi^2\ell^4 -2\xi\ell^4 r^2+12\xi^2\ell^6-2M\xi r^2\ell^4 +12M\xi^2\ell^6)}{\ell^2(r^2+2\xi\ell^2)^{3}}.
\end{equation}
\end{widetext}

\begin{figure*}[ht!]
	\centering
	\subfigure[]{\includegraphics[width=0.8\columnwidth]{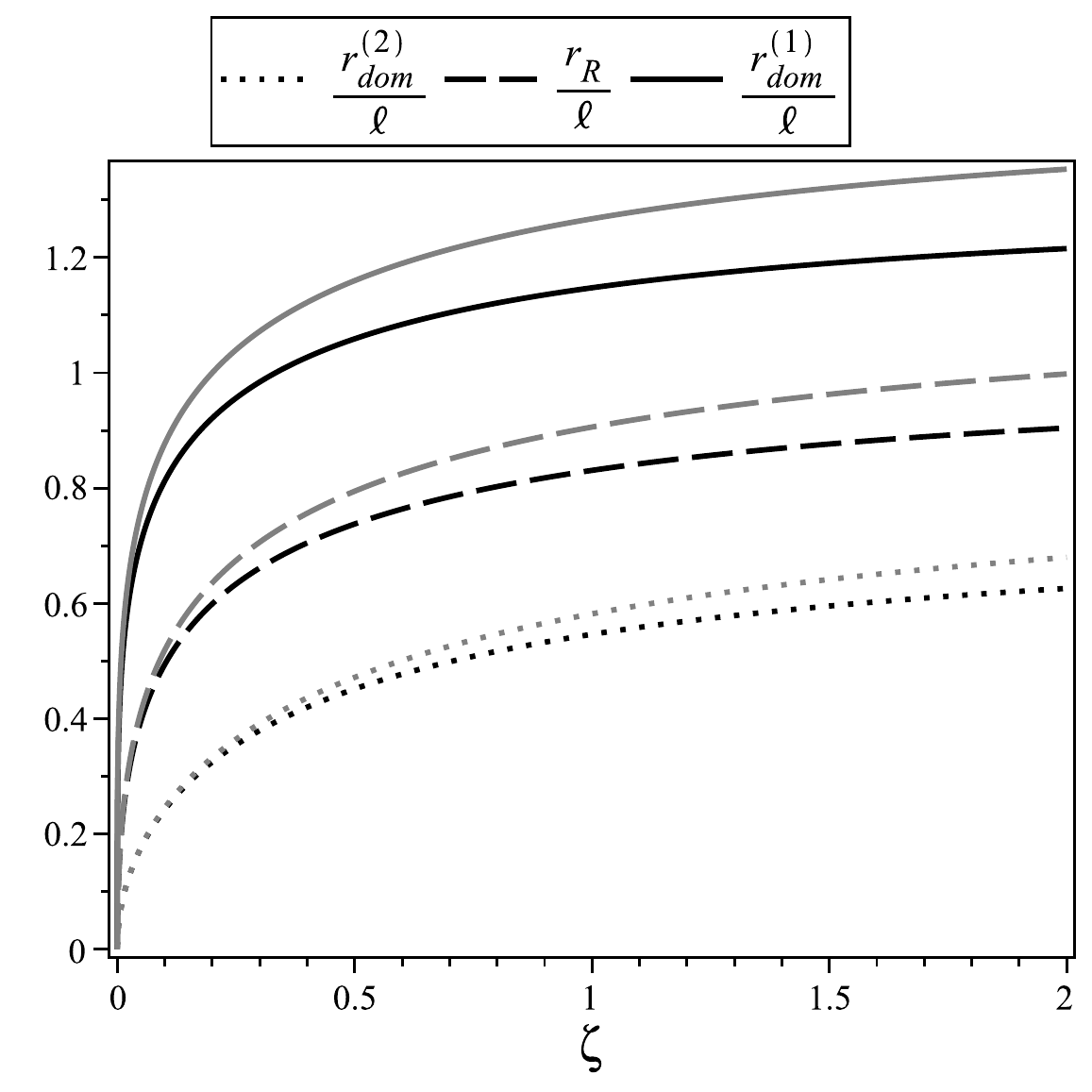}}~~~~~~~~~~
	\subfigure[]{\includegraphics[width=0.8\columnwidth]{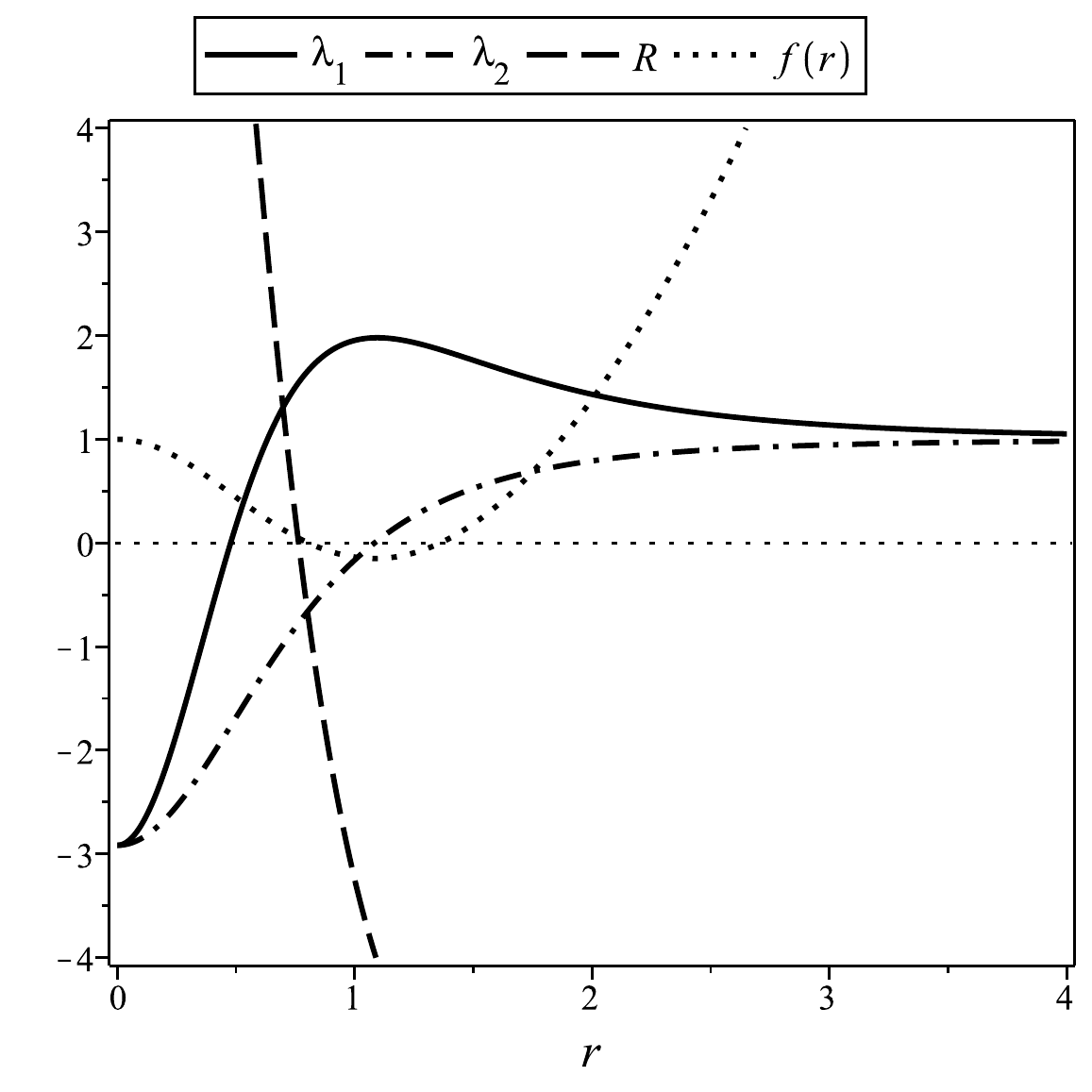}}
	\caption{Plots of ${r^{(1)}_{dom}/\ell}$, $r_{R}/\ell$ and ${r^{(2)}_{dom}/\ell}$ in terms of $\xi$ for $M=2.5,\textcolor{gray}{3.5}$ have been shown (left). Plots of the eigenvalues, ${\lambda_{1}}$, ${\lambda_{2}}$ and ${f(r)}$, $R$ for the third solution for $\ell=1,M=2.5,\xi=0.6$. }
	\label{fig3}
\end{figure*}

The Ricci scalar changes sign at
\begin{align}
r_{R}=\dfrac{\ell}{3}\sqrt{3X^{\frac{1}{3}}-\dfrac{6\xi}{X^{\frac{1}{3}}}(2\xi+M+1+3X^{\frac{1}{3}})},
\end{align}
where $X=144\xi^3+72(1+M)\xi^{2}+2\xi(1+M+2\xi)\sqrt{2\xi(1+M+650\xi)}$,  exhibiting repulsive gravity at distinct domains. As it can be seen from Fig. \eqref{fig3}a, for a constant value of $M$, $r^{(2)}_{dom}/\ell$ is larger than $r^{(1)}_{dom}/\ell$ and $r_{R}/\ell$. For $\xi=0$ and $M=-1$, all radii are equal. Thus, to have a black hole with $r_{rep}\geq r^{(1)}_{dom}$, the allowed region for $\xi$ is $(1+M)/6 \leq\xi$ and $\xi\leq M^{2}/8$. From the latter considerations, it can be concluded that $M$ lies inside $-1\leq M\leq-2/3$ or $M\geq 2$.

The behavior of the eigenvalues shows that the curvature becomes finite at the center of the source due to the presence of repulsive gravity. In addition, we also notice here the connection between repulsive gravity and the strong energy conditions.

\subsection{The fourth solution:
3D anti-de-Sitter with log correction and topological charge}

The fourth solution we consider here is \cite{He:2017ujy}
\begin{equation}
f(r)=-M+\dfrac{r^2}{\ell^2}-2q^2\left(\dfrac{q}{q+r}+\ln\left(\dfrac{q+r}{\ell}\right)\right) \ ,
\end{equation}
where $q$ and $M$ represent the electric charge and the ADM mass, respectively. Similar to the first case, this regular black hole  has a $\log$ term, and  is the solution of Einstein gravity coupled to non-linear electrodynamics ($L(r)=q^2(r-q)/(r+q)^3$). The non-linear electrodynamics also contains in this case the Maxwell theory limit in the weak field approximation and satisfies the weak energy condition.
The event horizon radii are given by
\begin{align}
r_{\pm}=\ell e^{\mathcal{B}}-q,
\end{align}
where $\mathcal{B}$ stands for the set of roots of the algebraic equation $2q^2\ell^2 Z e^{Z}+M\ell^2 e^{Z}+2q\ell e^{2Z}-q^2e^{Z}-\ell^2 e^{3Z}+2\ell q^3=0$.

The computation of the curvature eigenvalues $\lambda_{i}$, according to Eq.(\ref{eigenv}), leads to \cite{Luongo:2023aib}
\begin{align}
\lambda_{1}=&\dfrac{(1-\ell^2)q^3+rq^2(3+\ell^2)+3qr^2+r^3}{(q+r)^3\ell^2},\label{eqq3}\\
\lambda_{2}=&-\lambda_{3}=\dfrac{(1-\ell^2)q^2+2qr+r^2}{(q+r)^2\ell^2}.\label{eqq6}
\end{align}

\begin{figure*}[ht!]
	\centering
	\subfigure{\includegraphics[width=1\columnwidth]{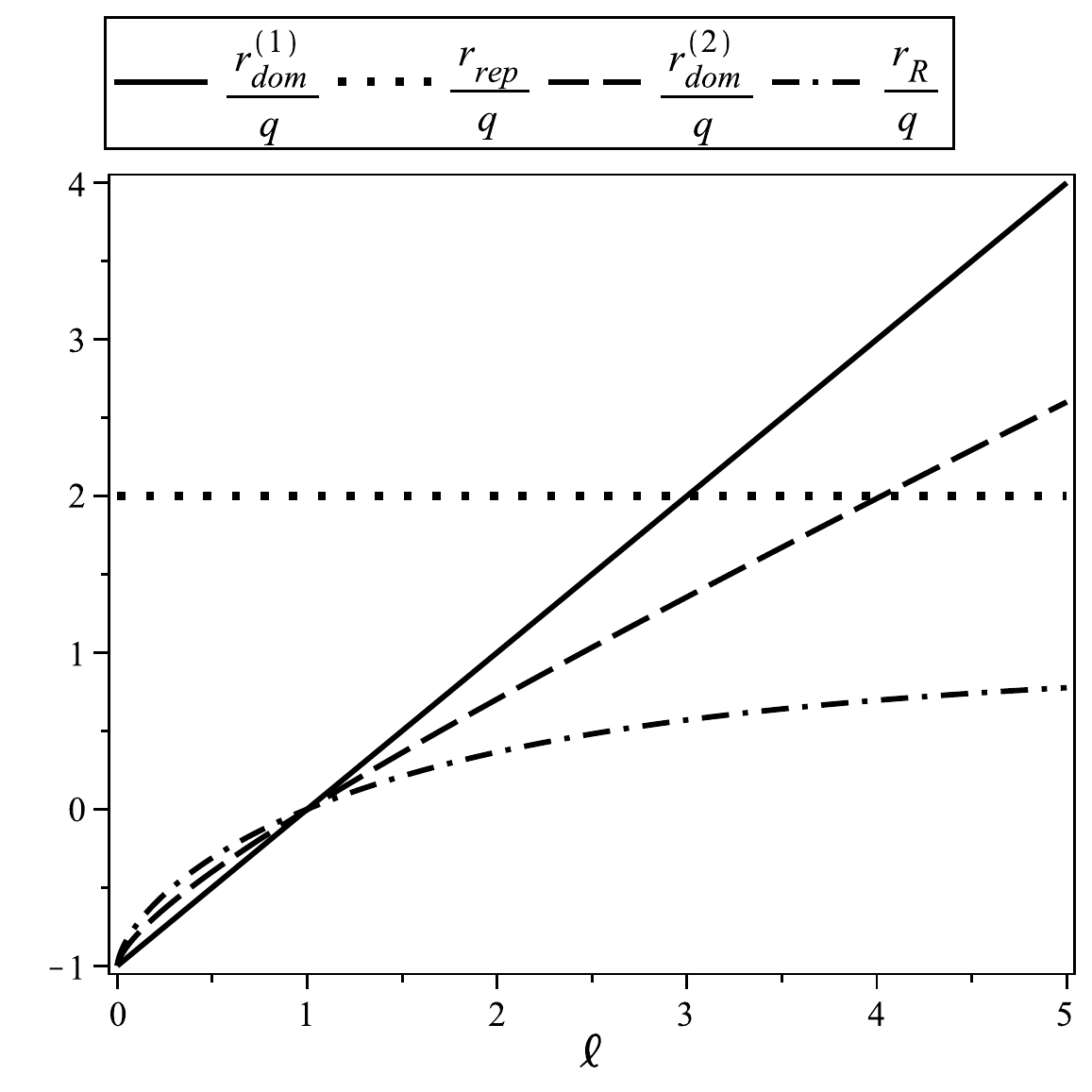}}~~~~~~~~~~
	\caption{Plots of ${r^{(1)}_{dom}/q}$, $r_{R}/q$, ${r^{(2)}_{dom}/q}$ and ${r_{rep}/q}$ in terms of $\ell$ have been shown. }
	\label{fig4}
\end{figure*}

For $r\to\infty$, the eigenvalues $\lambda_{i}$ go to $1/\ell^2$.
The first extremum that is reached when approaching from infinity is located at $r_{rep}=2q$, corresponding to a local maximum of $\lambda_{1}$, corresponding to the onset of repulsive gravity.
It is easy to see that $\lambda_{1}$ and $\lambda_{2}$ become zero at
\begin{align}
r^{(1)}_{dom} =&q(\ell-1),\\
r^{(2)}_{dom} =&\left((-\ell^2+\sqrt{\ell^{4}-\ell^{6}})^{\frac{1}{3}}+\dfrac{\ell^2}{(-\ell^2+\sqrt{\ell^{4}-\ell^{6}})^{\frac{1}{3}}}-1\right)q,
\end{align}
when approaching the source from infinity, for $0<\ell<1$ the first zero is in $r^{(2)}_{dom}$ and for $\ell>1$ the first zero is in $r^{(1)}_{dom}$, which is the point at which repulsive gravity becomes dominant. Further, the place where the strong energy condition is violated is equal to $r_{sec}=r_{rep}/2$. For $\ell\geq 3$, we have $r^{(1)}_{dom}\leq r_{rep}$.
On the other hand, evidence for repulsive gravity could be inferred also from the Ricci scalar, which is a first-order curvature invariant. In this case, one can obtain
\begin{equation}
R=2\dfrac{3(\ell^2-1)q^3+(\ell^2-9)rq^2-9qr^2-3r^3}{(q+r)^3\ell^2}.
\end{equation}
Moreover, the Ricci scalar for this metric changes sign at a given radius,
\begin{small}
\begin{align}
r_{R}=
\dfrac{q}{3}\ell^{\frac{2}{3}}\left[(9+\sqrt{81-\ell^2})^{\frac{1}{3}}+\dfrac{\ell^{\frac{2}{3}}}{(9+\sqrt{81-\ell^2})^{\frac{1}{3}}}\right]-q,
\end{align}
\end{small}
exhibiting repulsive gravity at distinct domains. As can be seen from Fig. \ref{fig4}a, $r^{(2)}_{dom}/\ell$ is larger than $r_{R}/\ell$ and $r^{(1)}_{dom}/\ell$.

\section{Theoretical discussion}\label{sezione4}

In the four-dimensional case, the most significant property of regular metrics is that repulsive gravity occurs being independent of linear combinations involving the free parameters on which the spacetimes depend. As a consequence, the value of mass does not affect the behavior of the regular eigenvalues, which is in contrast to what happens in the case of singular black hole solutions.
Thus, we find that there always exist specific masses that mimic the behaviors induced by repulsive gravity in regular solutions \cite{Luongo:2023aib}.

Here, in the first solution, we notice that the gravitational charge, $M$, does not appear in Eqs. \eqref{soluzionimetrica1}. This fact aligns with the outputs obtained for the repulsive radii ($r_{rep}=\sqrt{3}q$) that do not depend on $M$. This is due to the fact that the mass appears as an additive constant in the lapse function defined in Eq. \eqref{eqf2}. Consequently, we cannot conclude that mass can be used to mime gravitational effects in lieu of $q$ and $\ell$. The first solution appears, then, more physical with respect to the Bardeen spacetime in four dimensions.

The second and third solutions resemble the Hayward spacetimes. In this case, the gravitational charge plays the role of free background parameter and it appears manifestly in Eqs. \eqref{soluzionimodello2},  \eqref{repulsione2}, \eqref{soluzionimodello3} and  \eqref{repulsione3} as combination of the other constants. Surprisingly, $\ell$ for solution number three does not combine in Eq. \eqref{repulsione3}, showing a pathology in its choice. Consequently, solution number three appears less complete than the second solution. Together with solution one, it can be mimed by a singular solution that may show similar effects.

The last case appears similar to the first one. There, the value of mass is additive to the lapse function and consequently, it does not appear in the corresponding eigenvalues. This leads to similar considerations to the first case and, in fact, to characterize repulsive gravity one requires $q$ and $\ell$ only, regardless of the gravitational charge that, physically speaking, might determine the existence of the solution itself.

Since regular black holes are non-vacuum solutions of gravitational field equations, the sign change of the Ricci scalar can also be considered to determine repulsive gravity. For all four solutions, we obtained the locations, where the Ricci scalar changes sign. For the first and fourth solutions similar to the four-dimensional regular black hole, this radius is independent of the black hole mass, while for the second and third solutions, it depends on the black hole mass.

For all four spacetimes, the procedure and overall technique to obtain the domains of repulsion is similar to the standard four-dimensional case. Hence, the corresponding radii at which gravity changes sign appear similar in form to those inferred from the standard Bardeen and Hayward solution, albeit different in magnitude. Exceptional cases are those in which gravity may exhibit a change of sign exactly on the horizon, as in the first metric for example. This case appears particularly unclear as one approaching the horizon is expected to feel the action of gravity induced by the gravitational charge of the black hole itself, instead of being repulsed. The region appears therefore unphysical and aligns with the findings that seem to indicate the second spacetime as the most suitable one.

The overall treatment, however, points out severe physical differences among metrics. This may suggest that, although the net procedure provides similar mathematical results to the four-dimensional spacetime, 3D metrics can be quite different among them. Clearly, focusing on the second solution, one can imagine constructing more complicated 3D spacetime versions that can bypass the above problems, appearing more physical.

\section{Final outlooks and perspectives}\label{sezione5}

We here considered repulsive gravity in an invariant manner in the context of 3D regular spacetimes. Particularly, we investigated the effects of repulsive gravity comparing the Riemann curvature tensor expectations with respect to the eigenvalue approach that makes use of the bivector decomposition of the Riemann tensor.

To do so, we computed the corresponding first-order  invariant eigenvalues of the   four distinct metrics, entering two main classes, i.e., the first involving logarithmic corrections, and the second exhibiting different kinds of terms.

Specifically, we worked out those solutions inferred from non-linear electrodynamics and we analyzed local and asymptotic limits, showing the role played by an anti-de Sitter cosmological constant and how the topological charge influences the metrics.

Hence, our findings are based on two main 3D metrics, i.e., those being Bardeen-like, and those resembling Hayward-like behaviors. From each of them, we
investigated the sign of gravity checking the behavior of the eigenvalues, approaching from infinity the flat region placed at $r=0$.

The passage from attractive to repulsive gravity is also discussed in terms of the free parameters of the metrics themselves, emphasizing how and if those parameters are intertwined among them. An important conclusion of this analysis is that repulsive gravity can be interpreted as being responsible for the regular behavior of the solutions. In addition, we established that there is a connection between the radius of repulsion and the location where the strong energy condition is violated. This is certainly an unexpected result that is worth continuing to investigate in future works.

Particularly, once computed the onsets of repulsion, we reported the implications and physical consequences of our outcomes. Precisely, the second solution appeared more complete than the other three, exhibiting a mass and cosmological constant related to each other as well as it is found for standard singular solutions. In the same conclusion, it is not possible to arrive, considering the other three spacetimes, emphasizing the pathologies in their constructions.

In analogy to repulsion gravity for the four-dimensional case, we notice that the same mathematical form is found even for the 3D case.

Even though this supports the idea that the same procedure may be carried out, the above considerations on the eigenvalues suggest that future 3D metrics may be constructed assuming mass and free parameters, i.e., topological charge and/or cosmological constant, to be related among them in order to fix definite repulsive regions.

Future works will focus on how to compute repulsive regions in alternative solutions. Further, we will propose a more general treatment to get from a \emph{back-scattering procedure} a suitable choice of regular metric that resembles a black hole with no pathology in the repulsive regions. Moreover, a more detailed comparison between other classes of four and three-dimensional spacetimes will also be the subject of future efforts. Finally, for less dimensions than four, one may alternatively think to obtain the Cotton tensor eigenvalues to investigate repulsive regions. We will therefore investigate this occurrence to see whether new repulsive regions can be found adopting the Cotton tensor.

\section*{Acknowledgements}
The work of OL is partially financed by the Ministry of Education and Science of the Republic of Kazakhstan, Grant: IRN AP19680128. The work of HQ was partially supported by UNAM-DGAPA-PAPIIT, Grant No. 114520, and CONACYT-Mexico, Grant No. A1-S-31269.

\end{document}